\documentstyle[epsfig]{mn}
\newif\ifAMStwofonts

\AMStwofontstrue

\newcommand{\hst}{\textit{HST}}
\newcommand{\ngst}{\textit{NGST}}

\ifoldfss
  \newcommand{\rmn}[1] {{\rm #1}}

  \ifCUPmtlplainloaded \else
    \NewTextAlphabet{textbfit} {cmbxti10} {}
    \NewTextAlphabet{textbfss} {cmssbx10} {}
    \NewMathAlphabet{mathbfit} {cmbxti10} {} % for math mode
    \NewMathAlphabet{mathbfss} {cmssbx10} {} %  "   "    "
  \fi
  \ifAMStwofonts
    \ifCUPmtlplainloaded \else
      \NewSymbolFont{upmath} {eurm10}
      \NewSymbolFont{AMSa} {msam10}
      \NewMathSymbol{\upi}     {0}{upmath}{19}
      \NewMathSymbol{\umu}     {0}{upmath}{16}
      \NewMathSymbol{\upartial}{0}{upmath}{40}
      \NewMathSymbol{\leqslant}{3}{AMSa}{36}
      \NewMathSymbol{\geqslant}{3}{AMSa}{3E}

       \let\le=\leqslant
       \let\ge=\geqslant
    \fi
  \fi
\fi % End of OFSS

\ifnfssone
  \newmathalphabet{\mathit}
  \addtoversion{normal}{\mathit}{cmr}{m}{it}
  \addtoversion{bold}{\mathit}{cmr}{bx}{it}
  \newcommand{\rmn}[1] {\mathrm{#1}}

  \newmathalphabet{\mathbfit} % math mode version of \textbfit{..}
  \addtoversion{normal}{\mathbfit}{cmr}{bx}{it}
  \addtoversion{bold}{\mathbfit}{cmr}{bx}{it}
  \newmathalphabet{\mathbfss} % math mode version of \textbfss{..}
  \addtoversion{normal}{\mathbfss}{cmss}{bx}{n}
  \addtoversion{bold}{\mathbfss}{cmss}{bx}{n}
  \ifAMStwofonts
    \ifCUPmtlplainloaded \else
      \UseAMStwoboldmath
      \makeatletter
      \new@mathgroup\upmath@group
      \define@mathgroup\mv@normal\upmath@group{eur}{m}{n}
      \define@mathgroup\mv@bold\upmath@group{eur}{b}{n}
      \edef\UPM{\hexnumber\upmath@group}
      \new@mathgroup\amsa@group
      \define@mathgroup\mv@normal\amsa@group{msa}{m}{n}
      \define@mathgroup\mv@bold\amsa@group{msa}{m}{n}
      \edef\AMSa{\hexnumber\amsa@group}
      \makeatother
v      %
      \mathchardef\upi="0\UPM19
      \mathchardef\umu="0\UPM16
      \mathchardef\upartial="0\UPM40
      \mathchardef\leqslant="3\AMSa36
      \mathchardef\geqslant="3\AMSa3E

       \let\le=\leqslant
       \let\ge=\geqslant
    \fi
  \fi
\fi % End of NFSS release 1

\ifnfsstwo
  \newcommand{\rmn}[1] {\mathrm{#1}}

  \DeclareMathAlphabet{\mathbfit}{OT1}{cmr}{bx}{it}
  \SetMathAlphabet\mathbfit{bold}{OT1}{cmr}{bx}{it}
  \DeclareMathAlphabet{\mathbfss}{OT1}{cmss}{bx}{n}
  \SetMathAlphabet\mathbfss{bold}{OT1}{cmss}{bx}{n}
  \ifAMStwofonts
    \ifCUPmtlplainloaded \else
      \DeclareSymbolFont{UPM}{U}{eur}{m}{n}
      \SetSymbolFont{UPM}{bold}{U}{eur}{b}{n}
      \DeclareSymbolFont{AMSa}{U}{msa}{m}{n}
      \DeclareMathSymbol{\upi}{0}{UPM}{"19}
      \DeclareMathSymbol{\umu}{0}{UPM}{"16}
      \DeclareMathSymbol{\upartial}{0}{UPM}{"40}
      \DeclareMathSymbol{\leqslant}{3}{AMSa}{"36}
      \DeclareMathSymbol{\geqslant}{3}{AMSa}{"3E}

       \let\le=\leqslant
       \let\ge=\geqslant
    \fi
  \fi
\fi % End of NFSS release 2

\ifCUPmtlplainloaded \else
  \ifAMStwofonts \else % If no AMS fonts
    \def\upi{\pi}
    \def\umu{\mu}
    \def\upartial{\partial}
  \fi
\fi

\def\lsim{\mathrel{\rlap{\lower 3pt\hbox{$\mathchar"218$}}
     \raise 2.0pt\hbox{$\mathchar"13C$}}}
\def\gsim{\mathrel{\rlap{\lower 3pt\hbox{$\mathchar"218$}}
     \raise 2.0pt\hbox{$\mathchar"13E$}}}

\title[Gravitationally lensed distant SNe]
{A strategy for finding gravitationally-lensed distant supernovae}

\author[M. Sullivan et al.]{Mark~Sullivan,$^{1,2}$\thanks{E-mail: ms@ast.cam.ac.uk} Richard~Ellis,$^{2,1}$ Peter~Nugent,$^{3}$ Ian Smail,$^{4}$ \& 
Piero~Madau$^{1}$ \\
$^1$ Institute of Astronomy, Madingley Road, Cambridge CB3 OHA, UK\\
$^2$ California Institute of Technology, E. California Blvd, Pasadena CA 91125, USA\\
$^3$ E.O. Lawrence Berkeley National Laboratory, 1 Cyclotron Road, MS50-232, Berkeley, CA 94720, USA \\
$^4$ Department of Physics, University of Durham, South Road, Durham DH1 3LE, UK\\
}

\date {Accepted ---. Received ---; in original form ---.}
\begin{document}
\label{firstpage}
\maketitle

\begin{abstract}
  Distant Type Ia and II supernovae (SNe) can serve as valuable probes
  of the history of the cosmic expansion and star formation, and
  provide important information on their progenitor models.  At
  present, however, there are few observational constraints on the
  abundance of SNe at high redshifts.  A major science driver for the
  \textit{Next Generation Space Telescope (NGST)} is the study of such
  very distant supernovae. In this paper we discuss strategies for
  finding and counting distant SNe by using repeat imaging of
  super-critical intermediate redshift clusters whose mass
  distributions are well-constrained via modelling of strongly-lensed
  features.  For a variety of different models for the star formation
  history and supernova progenitors, we estimate the likelihood of
  detecting lensed SNe as a function of their redshift. In the case of
  a survey conducted with \hst, we predict a high probability of
  seeing a supernova in a single return visit with either WFPC-2 or
  ACS, and a much higher probability of detecting examples with $z>1$
  in the lensed case. Most events would represent magnified SNe\,II at
  $z\simeq1$, and a fraction will be more distant examples. We discuss
  various ways to classify such events using ground-based infrared
  photometry. We demonstrate an application of the method using the
  \hst\ archival data and discuss the case of a possible event found
  in the rich cluster AC\,114 ($z=0.31$).
\end{abstract}


\section{Introduction}

Recent observational efforts to detect high-redshift ($z$) supernovae
(SNe) have demonstrated their value as cosmological probes.  The
discovery and systematic study of faint, distant type Ia supernovae
(SNe\,Ia) has led to renewed progress in constraining the cosmic
expansion history \cite{riess98,perlmutter99}. Two independent
corrected Hubble diagrams based on SNe\,Ia discovered via the
Supernovae Cosmology Project (Perlmutter et~al.~1999) and the
High-Redshift Supernovae Search Team (Riess et al.~1998) yield a trend
that strongly excludes the hitherto popular Einstein de Sitter
universe and, for a spatially flat inflationary universe consistent
with recent microwave background measurements
\cite{hancock98,debernardis00}, suggests a significant non-zero
cosmological constant, $\Lambda\simeq0.7$ \cite{perlmutter99,riess98}.

Such studies have now identified over 100 $z>0.2$ SNe\,Ia, and, using
controlled subsets of these SNe\,Ia samples, the first constraints are
now emerging on the rate of their occurrence (Pain et al.\ 1996, 2000a).
The observational determination of these rates is also important in
many cosmological applications, for example as a diagnostic of the
cosmic star formation history (SFH) and metal enrichment at high-$z$.
SNe are independent of some of the biases associated with traditional
tests based on flux-limited galaxy samples. Such biases are
increasingly of concern given the steep luminosity functions now being
witnessed for star-forming galaxies at all redshifts
\cite{steidel99,sullivan00} and, though there remain uncertainties in
the correct treatment of dust in high-$z$ type II supernovae
(SNe\,II), they offer an alternative approach to studying the cosmic
SFH.

SNe\,II are caused by the catastrophic core collapse of massive
($>8\,\rmn{M}_{\odot}$) stars and, unlike SNe\,Ia, represent a direct
probe of the massive SFH due to their short progenitor lifetimes --
the evolution of the SNe\,II rate should closely match that of the
cosmic SFH \cite{madau98,sadat98}.  Unfortunately, due to their
intrinsically faint, heterogeneous nature, and their likely occurrence
in dusty starburst regions, to date they have yet to be exploited in
cosmological studies. However, as the predicted SNe\,II counts to
faint survey limits (e.g. $I=26$) will outnumber those of SNe\,Ia
(Madau et al.~1998; Dahlen \& Fransson~1999, Porciani \& Madau~2000),
the detection of SNe\,II at higher-$z$ should become more practical as
these fainter limits are attained -- either through facilities such as
the \ngst\ \cite{dahlen98} or via studies of the type introduced here.

SNe\,Ia, by contrast, arise from lower mass ($\simeq
3-8\,\rmn{M}_{\odot}$) stars, and their rate depends strongly on the
(as yet) unknown configuration of the progenitor system as well as the
cosmic SFH. The SNe\,Ia rate therefore cannot yet be used to directly
probe the SFH without a more detailed knowledge of the physics of
SNe\,Ia explosions.  The observed rate does however offer important
implications for the possible progenitor models
\cite{madau98,yungelson98} and, at $z>1.5$, the predicted
metal-dependency of the SNe\,Ia rate will provide valuable clues to
the metal enrichment at high-$z$ \cite{kobayashi98}, as well as for
the practicality of extending the Hubble diagram for more precise
cosmological tests \cite{ruiz98}.

The study of distant SNe of both types is one of the key programs
underpinning the \ngst\ \cite{dahlen98}. The optimum strategy for
finding and studying SNe, particularly with respect to the
investigation of SNe\,II and the SFH at $z>1$, clearly depends on
their abundance. In this paper, we suggest a strategy for producing
the first observational constraints using existing facilities. The
strategy proposed follows the idea proposed by Kolatt \&
Bartelmann~(1998) and a similar approach used to locate faint sub-mm
sources \cite{smail97a}. By exploiting the magnification bias afforded
by counting sources viewed through the cores of strong lensing
clusters, fainter detection thresholds can be reached
\cite{blain99a}. In the case of the sub-mm surveys the intrinsic flux
limits achieved are fainter than those reached in more ambitious
dedicated deep surveys. Adopting the same principle, optical searches
for lensed SNe should provide a first glimpse of the evolution of SNe
at high-$z$ ahead of the launch of \ngst.

A plan of the paper follows. In $\S$2, we review the theoretical
aspects of gravitational lensing as applicable to our proposed search
and discuss the uncertainties associated with modelling SNe rates.
This allows us to predict the source counts and redshift distribution
for a survey with and without the effect of lensing. In $\S$3 we
discuss a possible application of the technique using the \hst\ data
archive. Only two rich clusters with the necessary lensing credentials
have been multiply-visited in a single passband by \hst, and we
discuss one possible SNe found in each. In $\S$4 we discuss the
possibilities of classifying such detections via various follow-up
strategies. We present our conclusions in $\S$5.  Throughout the
paper, for consistency with the SNe\,Ia and microwave background
results, we assume a lambda dominated cosmology of $\Omega_{M}=0.3$,
$\Omega_{\Lambda}=0.7$, and $h=0.5$, where
$H_0=100\,h\,\rmn{km\,s}^{-1}\rmn{\,Mpc}^{-3}$.

\section{PREDICTED SUPERNOVA COUNTS}

\subsection{Introduction}

The approach proposed is identical to that successfully employed in
probing the faint sub-mm sources \cite{smail97a,blain97}. The core of
a rich lensing cluster magnifies suitably-located background sources
by factors of several. The critical difference with the sub-mm studies
is that, for faint SNe studied at optical wavelengths, the exquisite
angular resolution of \hst\ is essential. The modest boost in
detectability afforded by surveying through a massive foreground
cluster significantly extends the mean detectable redshift. We
primarily consider a search using the \hst\ and the Wide Field
Planetary Camera 2 (WFPC-2).  The high spatial resolution, dark sky
background and relatively large field-of-view (5.71\,$\rmn{arcmin}^2$)
of this telescope/instrument combination ensures faint survey limits
can be reached in a reasonable exposure time. We also briefly consider
the advantages of using the forthcoming Advanced Camera for Surveys
(ACS) \hst\ instrument, with a larger (11.33\,$\rmn{arcmin}^2$)
field-of-view.

In the rest of this section we investigate this observational approach
by calculating the number density $N(m,z)$ of SNe of different types
with and without the magnification produced by a massive cluster
located at various redshifts. Below we introduce the theoretical
aspects of both the lensing and the intrinsic SNe properties as a
function of redshift.
 
\subsection{Cluster Mass Models}
\label{modelclusters}

The role of clusters in gravitational lensing has been reviewed by
many authors (see, for example, Blandford \&
Narayan~\shortcite{blandford92} or Hattori, Kneib \&
Makino~\shortcite{hattori99} and references therein). A gravitational
lens will distort and magnify a background source by an amount that
depends on both the position of the source behind the cluster, and the
relative redshifts of the source and the lens. This results in
different background regions suffering different degrees of
amplification, depending on their location behind the cluster. Through
the location and spectroscopic identifications of multiply-imaged arcs
\cite{mellier99}, it is possible to correct for this effect by
utilising a well-constrained cluster mass model to the accuracy
required. These models involve multiple component mass distributions
that are able to trace the potential well of a cluster and its more
massive individual member galaxies \cite{kneib96}. The uncertainties
in these best cases is currently $\simeq 10-20$ per cent
\cite{blain99a}.

The degree of magnification that one can expect in the central regions
of such massive clusters -- the areas sampled when conducting lensed
SNe searches -- are typically of order $\mu\simeq3$--4 for sources at
$z_s=2$ \cite{blain99a}, where $\mu$ is defined as the ratio of the
magnified flux to the flux measured in the absence of lensing.
However, this boost in sensitivity -- or effective increase in the SNe
number density as fainter SNe become brighter than the survey limiting
magnitude due to amplification -- is offset by the competing effect of
depletion. This results in a decrease in the SNe number density, as
the unit solid angle in the unlensed source plane \textit{increases}
by a factor of $1/\mu(z)$ in the lensed case, resulting in a
\textit{decrease} in the SNe number density by $1/\mu(z)$.

When attempting to determine a precise count rate of lensed sources,
accurate modelling is essential. Here, we choose to model the cluster
mass distribution more simply via a singular isothermal sphere (SIS),
where the mass density $\rho$ of the virialized dark matter halo of a
cluster is given by $\rho(r)=\sigma_v^2/2 \pi G r^2$, where $r$ is a
radial coordinate, and $\rho$ is parameterized through $\sigma_v$, the
one-dimensional velocity dispersion. The mass $M$ of a halo interior
to $r$ is then $M(r)=2\sigma_v^2r/G$. This corresponds to a constant
deflection angle for incident light rays, $\beta=4 \pi (\sigma_v/c)^2
=28.8'' \;(\sigma_v/1000\; {\rm km}\;{\rm s}^{-1})^2$, always pointing
towards the lens centre of symmetry.  For a source that, in the
absence of lensing, would be seen at an angular distance $\theta$ from
this centre, the lens equation leads to an image at $\theta_+=\theta+
\theta_E$, with magnification $\mu_+=\theta_E/\theta+1$, where
$\theta_E=\beta D_{ls} /D_{os}$ is the Einstein radius, and $D_{os}$
and $D_{ls}$ are the angular diameter distances between the observer,
the lens, and the source.  If the alignment is $\theta\le \theta_E$
(strong lensing), a second image is produced at
$\theta_-=\theta-\theta_E$ with magnification
$\mu_-=\theta_E/\theta-1$.

High-resolution $N$--body simulations have recently shown, however,
that haloes formed through hierarchical clustering have a universal
density profile which is shallower than isothermal (but still
diverging like $\rho\propto r^{-1}$) near the halo centre, and steeper
than isothermal (with $\rho\propto r^{-3}$) in its outer regions
(Navarro, Frenk, \& White 1997, hereafter NFW). This profile takes the
form:

\begin{equation}
\rho(r)=\frac{\rho_c\,\delta_c}
{(r/r_s)(1+r/r_s)^2} 
\end{equation}

\noindent
where $\rho_c=3\rmn{H}^2(z)/8\pi\rmn{G}$ is the critical density at
the lens redshift, $\delta_c$ is a characteristic over-density, and
$r_s=r_{200}/c$ is a scale radius, where $c$ is a dimensionless number
measuring the ``concentration'' of the halo, and the virial radius,
$r_{200}$, is defined as the radius inside which the mass density is
equal to $200\rho_c$. $\delta_c$ is related to $c$ via

\begin{equation}
\delta_c=\frac{200}{3}\frac{c^3}{\ln{(1+c)}-\frac{c}{1+c}}
\end{equation}

\noindent
We can then see that $M(r_{200})=800\pi\rho_cr^3_{200}/3$. The NFW
haloes are thus defined by two parameters; $c$, and either $r_{200}$ or
$M_{200}$.

The lens equation for the NFW density distribution has been computed
by several authors \cite{bartelmann96,maoz97,wright99}, and a good
summary of its application can be found in Hattori et
al.~\shortcite{hattori99}. With respect to SIS profiles containing the
same total mass and located at the same redshift, NFW lenses have
smaller cross sections for producing multiple images, but this is
compensated for by the fact that the average magnification of the
images is higher for NFW haloes.  In other words, NFW haloes are
efficient magnifiers but poor image-splitters -- a SIS profile
underestimates the magnification obtained compared to an NFW profile.

In the following calculations, we model the lensing clusters using
both SIS and NFW profiles. As a guide to the benefits of searching for
SNe through clusters, we have explored two hypothetical cases: a
survey conducted through a sample of (known)
$\sigma_v=1000\rmn{\,km\,s}^{-1}$ $z\simeq0.2$ clusters, and a similar
sample of $z\simeq0.6$ clusters now being located through the MACS
Survey (see for example Ebeling, Edge \& Henry~2000). To meaningfully
compare the magnifications of SIS and NFW profiles, we assume that for
both profiles the clusters have identical masses interior to the
virial radius, $r_{200}$. We then equate the expressions for $M(r)$ at
$r_{200}$ for both profiles, and hence set $r_{200}$ for the NFW
profile given a value for $\sigma_v$. For the NFW profiles, the
concentration parameter $c$ for each cluster is calculated for our
cosmology using the FORTRAN routine \textit{charden.f} provided by J.
Navarro.

\begin{figure}
  \epsfig{figure=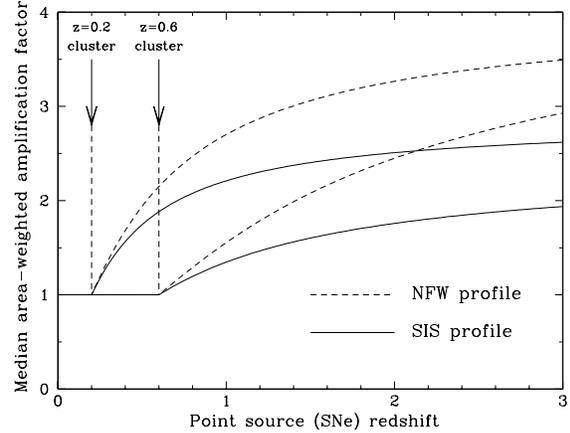,angle=270,width=80mm}
  \caption{The median area-weighted magnification suffered by a point 
    source as a function of redshift behind typical
    $\sigma_v=1000\rmn{\,km\,s}^{-1}$, $z=0.2$ and $z=0.6$ clusters,
    for a search conducted using WFPC-2. The amplifications expected for both
    NFW and SIS profiles are shown.}
  \label{amplificationfigure}
\end{figure}

The median area-weighted amplifications are calculated by considering
a pixel matrix corresponding to the WFPC-2 CCD, and determining the
amplification suffered by a background source were it to appear in
each element. The median pixel amplification can then be found.
Fig.~\ref{amplificationfigure} shows this amplification of a
background source as a function of source redshift for two typical
clusters. Typical median amplifications of $\mu\simeq2$--3 are typical
of $z\ge1.5$ sources, corresponding to a brightening in magnitudes of
$\,\simeq1$.

\subsection{Supernova Rates}
\label{modelsne}

We now estimate the number of SNe we would expect to find to a given
survey magnitude limit at a given epoch, in the case of a blank field
(i.e. without any significant lensing magnification) and through a
rich cluster. We will not include those (unlensed) SNe that occur in
the cluster galaxies, and will therefore underestimate the total
number of new SNe that will be found in a lensed search. The number
counts depend on both the intrinsic redshift-dependent SNe rate, and
that period of the light curve during which time the SNe appears
brighter than the survey search limit.

A number of workers have modelled the expected evolution in the SNe
rate with redshift. Most have followed a similar route to that
employed by Madau et al.~\shortcite{madau98}, who base their estimates
on the observed luminosity densities which trace the cosmic star
formation history (SFH). Using various diagnostics, the cosmic SFH can
now be traced to high redshift ($z\simeq 4$), although some details
remain controversial. As the progenitor lifetimes for SNe\,II are very
short by cosmological standards, the SNe\,II rate can be calculated
directly from the SFH provided an initial mass function (IMF) is
assumed. Here we adopt a Salpeter~\shortcite{salpeter} IMF with lower
and upper mass cut-offs of $0.1\,\rmn{M}_{\sun}$ and
$125\,\rmn{M}_{\sun}$ respectively.

In modelling the SNe\,II rate, the form of the cosmic SFH is critical.
The original study of the evolution in the comoving star-formation
rate based on optically selected galaxy surveys \cite{madau96}
suggested a sharp rise from $z=0$ to a peak at $z\simeq 1$--2 by a
factor of almost 10 \cite{lilly96,madau96,connolly97}, followed by a
decline at high-$z$. However, more recent studies suggest that the
evolution of the SFH up to $z\simeq 1.5$ may have been overestimated,
perhaps only increasing by a factor of $\simeq 4$
\cite{tresse98,cowie99,sullivan00}, and that the SFR at high-$z$ may
be obscured due to large amounts of dust extinction (see, for example,
Pettini et al.~1998 and Blain et al.~1999b).  The form
of this evolution to $z\simeq2$ is crucial for our studies when
predicting the SNe\,II rate.

We consider two illustrative SFHs. The first (SFH-I) is taken from Madau \&
Pozetti~\shortcite{madau00}, and provides a convenient analytical fit
of the form:

\begin{equation}
\dot\rho_s(z)={0.23\,e^{3.4z}\over e^{3.8z}+44.7},
\end{equation}

\noindent
which assumes an IMF lower mass cut-off of $\simeq0.5\,\rmn{M}_{\sun}$
and an Einstein de Sitter (EdS) Universe. Multiplying by a factor of
1.67 will convert to a Salpeter IMF with a lower mass cut-off of
$0.1\,\rmn{M}_{\sun}$. We convert the SFH to the $\Lambda$-dominated
universe assumed here by computing the difference in luminosity
density between an EdS Universe and our cosmology, and applying this
correction to the fit above. The SFH fit matches most UV-continuum and
H$\alpha$ luminosity densities from $z=0$ to $z=4$, includes a mild
correction for dust of $A_{1500}=1.2$ mag ($A_{2800}=0.55$ mag), and
implies a local SNe\,II rate of
$6.2\times10^{-5}\,\rmn{SNe\,yr^{-1}\,Mpc}^3$, very close to that
observed by Cappellaro, Evans \& Turatto~\shortcite{cappellaro99}.

\begin{figure}
  \epsfig{figure=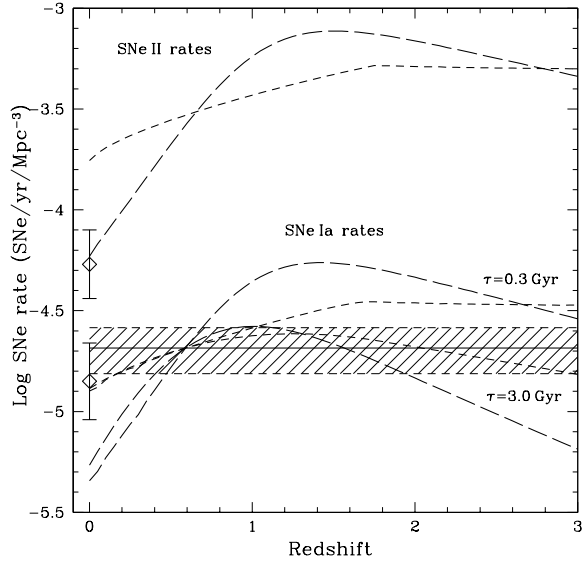,width=80mm}
  \caption{The adopted SNe rates as a function of redshift for both
    SNe\,II (top) and SNe\,Ia (bottom). The long dashed lines refer to
    rates derived using SFH-I, the short dashed lines refer to SFH-II.
    The SNe\,Ia rates are shown using two `delay' times ($\tau=0.3$ \&
    $3.0\,\rmn{Gyr}$). The horizontal solid line and shaded zone
    indicates the $z=0.55$ SNe\,Ia rate and its uncertainty obtained
    by Pain~et~al.~(2000a), used to calculate the efficiencies of the
    SNe\,Ia rates. Local SNe rates determined by Cappellaro et
    al. (1999) are shown for comparison.}
  \label{snerates}
\end{figure}

However, the SFR evolution in this model to $z\simeq1.5$ is both
stronger, and results in a lower local SFR, than more recent
measurements (Cowie et al.~1999; Sullivan et al.~2000). Accordingly,
we also consider a second SFH (SFH-II) with a shallower evolution (a
factor of $\simeq4$ from $z=0$ to $z\simeq1.75$ in an EdS Universe,
and constant thereafter), but one which over-produces the local
SNe\,II rate. The SNe\,II rates as a function of redshift are shown in
Fig.~\ref{snerates}.

The SNe\,Ia rate is more difficult to model due to the uncertain
nature of the progenitor system. The evolution expected depends
critically on whether SNe\,Ia occur in double or single degenerate
progenitor systems (for a review see Nomoto
et~al.~\shortcite{nomoto99} and references therein), and there is much
debate in the literature as to what form any evolution should take. We
consider a commonly used empirical approach, and represent the
uncertainty in terms of two parameters (Madau et al.~1998; Dahlen \&
Fransson~1999). The first is a delay time $\tau$, between the binary
system formation and SNe explosion epochs, which defines a
(time-independent) explosion probability per white dwarf. The second
is an explosion efficiency, $\eta$, which accounts for the fraction of
binary systems that never result in a SNe. This efficiency $\eta$ is
constrained using a preliminary determination of the SNe\,Ia rate at
$z=0.55$ of $1.66\pm0.42\,h^{3}10^{-4}\rmn{Mpc}^{-3}\rmn{yr}^{-1}$ by
Pain et~al.~\shortcite{pain00a} using a sample of 38 SNe\,Ia from the
SCP (see also Pain et al.~2000b).

We consider two SNe\,Ia evolutions, shown in Fig.~\ref{snerates}.  The
first has a small value of $\tau=0.3\,\rmn{Gyr}$ (corresponding to a
shallower decline at high-$z$,) whilst the second is a larger value of
$\tau=3.0\,\rmn{Gyr}$, which produces a steeper drop-off. For
comparison, the $z=0.55$ rate of Pain et~al.~\shortcite{pain00a} is also
shown in Fig.~\ref{snerates}, together with the difference that arises
from using the two SFHs, and how these histories tend to over or
under-predict the local SNe rates.

While much published data exists for SNe\,Ia, by contrast there is a
dearth of data for SNe\,II and a great variety in their spectra, peak
magnitudes and light curves. For the purposes of this paper, we adopt
the models of Gilliland, Nugent \& Phillips~\shortcite{gilliland99}
(hereafter GNP) which incorporate recent light curve templates,
$k$-corrections and luminosity functions for SNe\,Ia and II. These
models include observed dispersions in the peak magnitudes of SNe\,Ia
and II, which are included in our simulations of our SNe counts (see
GNP for a full discussion of the models and simulations).

The major advantage to using GNP's models lies in the fact that the
Spectral Energy Distributions (SEDs) are more realistic than those
employed previously for rate calculations and follow the true temporal
evolution of these objects. The SNe\,Ia template was constructed using
data from \textit{IUE} and \hst\ \cite{iuesne,kir92a} in the UV and a
large number of ground-based observations in the optical. Spectrum
synthesis calculations were carried out by PN for the SNe\,II's using
the non-LTE code PHOENIX 9.0 \cite{hb971,hb972}. The spectra of
SNe\,1979C \cite{iuesne}, 1992H and 1993W (courtesy D. Leonard) were
fit at several epochs. The resultant fits of these spectra provided a
nice match to the observed data and yield a crude SED template for the
SNe\,II's. The major emphasis of this effort was to model the UV as
well as possible (this portion of the SED dominates in high-$z$
searches using filters bluer than $1\mu\rmn{m}$), since the simple
assumption of blackbody SEDs has been shown to be quite inaccurate
(see fig.~10 of GNP).

\subsection{Results}
\label{predictedcounts}

We can now combine the two mass models from $\S$\ref{modelclusters} and the
various SNe rate models from $\S$\ref{modelsne} to estimate the number
counts of SNe with and without the presence of a strong-lensing
cluster.

As with blank-field surveys, searches such as those outlined here
require two images of the same rich cluster, but with the two epochs
of observation separated by a suitable amount to allow the SNe to
become visible.  When estimating the simple count rate of lensed
high-$z$ SNe, therefore, both forward (SNe in second image) and
backward (SNe in first image) searches can be included, and to allow
for the time dilation of the high-$z$ SNe, we assume a time of at
least 2\,yrs between the two exposures.  Table~\ref{count_table} gives
the counts expected to survey limits of $I_{814}=26.0$ \& $27.0$,
assuming $z_{\rmn{cluster}}=0.2\,\&\,0.6$.

When comparing lensed and unlensed counts, one of the key factors is
the slope of the SNe number counts, $\alpha=\rmn{d}\log N/\rmn{d}m$.
For $\alpha>0.4$, we expect lensing to increase the number counts,
while for $\alpha<0.4$, lensing will have the opposite effect
(depletion). The slopes in our models depend on i) the intrinsic SNe
properties, contained in the models of GNP, and ii) our assumed
redshift evolution of the SNe number density. The models of GNP
include the best-determined uncertainties in the SNe properties, for
example in the dispersion of the peak SNe magnitude, and our
simulations therefore account for this first uncertainty. The redshift
evolution of the SNe numbers depends on which SFH we use, and, for
SNe\,Ia, the value of the delay time, $\tau$. By running simulations
for the unlensed case, we find that typical values of $\alpha$ for
SNe\,II for SFH-I are $\alpha\simeq0.35$, and for SFH-II
$\alpha\simeq0.32$. There is little variation in these values between
simulation runs.

\begin{figure}
\epsfig{figure=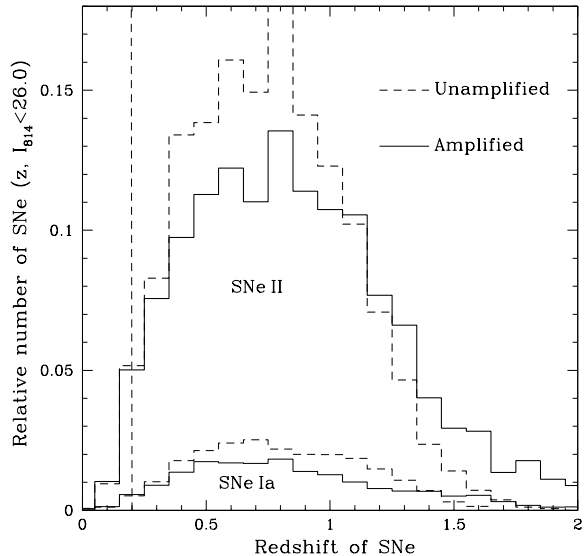,width=80mm}
\caption{Redshift distribution of a search for lensed SNe generated
  from a Monte-Carlo simulation using the models of $\S 2.2 \& 2.3$,
  to the limit $I_{814}=26.0$. Both SNe\,II (heavy lines) and SNe\,Ia
  (light lines), magnified (solid line) and unmagnified (dashed line),
  are shown. The position of the (NFW profile) lensing cluster is
  shown by the dashed line at $z=0.2$. Lensing alters the redshift
  distribution, resulting in the detection of a greater proportion of
  high-redshift SNe of both types. Only forward searches are included
  here.}
\label{magnz}
\end{figure}

The number-redshift relation is plotted in Fig.~\ref{magnz}. As one
would expect, the figure shows there is little difference in the rate
of SNe up to $z_{\rmn{cluster}}$. From $z_{\rmn{cluster}}$ to
$z\simeq1$, the lensed counts are lower than those appropriate for
blank field searches. This is due to the effect of depletion mentioned
above, which unfortunately negates the lensing benefits. However, the
important differences begin at $z>1$, where we sample a fainter, more
numerous population of SNe. Lensing has the principal effect of
increasing the proportion of SNe found at high-$z$, bringing fainter,
magnified examples into view.  There is then a corresponding extension
in the redshift distribution. For example, for the $z=0.2$ NFW
cluster, the mean redshift of detection changes from $z\simeq0.70$ to
$z\simeq0.85$, and the number of detections at $z>1$ increases from 20
to 31 per cent.

The numbers drop sharply by $z=2$, even in the lensed case, due to the
UV drop-off in the SNe spectrum being redshifted into the $I_{814W}$
filter, which prevents the detection of higher-$z$ SNe. We note that
the use of \hst\ and an infra-red search would allow the detection of
lensed $z>2$ SNe which would be undetectable in the unlensed case.

\begin{table*}

\caption{The predicted number of SNe to survey limits of $I_{814}=26.0$ \& $27.0$
 per WFPC-2 field for both a blank field and for a lensed survey (assuming
 $z_{\rmn{lens}}=0.2/0.6$ and the two lensing models). In each column the main
 entries refer to SFH-I and the values in parenthesis to SFH-II. The totals are
 the sum of IIL, IIP and a mean of the two Ia values. The bottom two lines show
 the expected counts for a survey using ACS and one of the scenarios. Also shown
 are the percentage of SNe found at $z>1.0$ \& $z>1.5$. Both forward and backward
 searches are included.}

\begin{center}
\begin{tabular}{cccccccc} \hline 
Survey &Lensing & \multicolumn{2}{c}{\textit{SNe\,Ia}} & \textit{SNe\,II} & Totals & \multicolumn{2}{c}{Fraction (\%) of SNe\,II with:}\\
Limit &Model & $\tau=0.3$ & $\tau=3.0$ & (IIL+IIP) & & \textit{$z>1.0$} & \textit{$z>1.5$}\\
\hline
$I=26.0$, \textbf{WFPC-2} & Unlensed & 0.32 (0.25) & 0.22 (0.23) & 1.44 (1.68) & \textbf{1.71 (1.92)}& 23.2 (14.4) & 1.4 (1.2)\\
&SIS, $z=0.2$ & 0.25 (0.19) & 0.18 (0.18) & 1.30 (1.44) & \textbf{1.52 (1.63)}& 30.3 (20.1) & 4.0 (3.3)\\
&SIS, $z=0.6$ & 0.31 (0.24) & 0.22 (0.22) & 1.47 (1.68) & \textbf{1.74 (1.91)}& 26.4 (16.1) & 2.6 (1.8)\\
&NFW, $z=0.2$ & 0.23 (0.18) & 0.16 (0.17) & 1.25 (1.40) & \textbf{1.45 (1.58)}& 31.0 (20.8) & 4.6 (3.8)\\
&NFW, $z=0.6$ & 0.30 (0.23) & 0.21 (0.21) & 1.46 (1.69) & \textbf{1.72 (1.91)}& 26.9 (17.5) & 3.2 (2.5)\\
$I=27.0$, \textbf{WFPC-2} & Unlensed & 0.56 (0.41) & 0.38 (0.37) & 2.86 (3.10) & \textbf{3.33 (3.49)}& 32.7 (22.6) & 3.8 (2.8)\\
&SIS, $z=0.2$ & 0.42 (0.30) & 0.27 (0.27) & 2.34 (2.42) & \textbf{2.69 (2.71)}& 37.7 (26.8) & 6.4 (5.9)\\
&SIS, $z=0.6$ & 0.51 (0.38) & 0.36 (0.34) & 2.85 (3.05) & \textbf{3.29 (3.41)}& 34.5 (23.2) & 5.4 (4.2)\\
&NFW, $z=0.2$ & 0.38 (0.29) & 0.25 (0.25) & 2.20 (2.31) & \textbf{2.52 (2.58)}& 37.9 (27.1) & 7.1 (6.3)\\
&NFW, $z=0.6$ & 0.49 (0.37) & 0.33 (0.33) & 2.77 (2.97) & \textbf{3.18 (3.34)}& 33.5 (23.4) & 5.4 (4.7)\\
$I=26.0$, \textbf{ACS}& Unlensed & 0.63 (0.50) & 0.44 (0.46) & 2.86 (3.33) & \textbf{3.39 (3.83)}& 23.2 (14.4) & 1.4 (1.2)\\
&SIS, $z=0.2$ & --- (0.40) & --- (---) & --- (2.92) & \textbf{--- (3.32)}& --- (19.7) & --- (3.0)\\
\hline

\end{tabular}
\end{center}
\label{count_table}
\end{table*}

\section{Application to the \hst\ Archive}
\label{detection}

Clearly, the number of detectable SNe depends on many variables which
forms a strong motivation for this search, particularly in view of the
need to quantify the high redshift rate in anticipation of detailed
studies with \ngst. A robust prediction from Table~\ref{count_table}
is the high chance of discovering a new SNe during a repeat exposure
of a previously studied lensing cluster.  Specifically, to a
reasonably modest magnitude limit (achievable in $\simeq1$--2 orbits),
we can expect to discover 1 to 3 SNe per visit if we combine forward
and backward searches.

To demonstrate the potential of such a search, we have analysed the
\hst\ data archive in the case of repeated observations of
strong-lensing clusters for which mass models are available.  Though
there are few such repeated visits with the same filters, exposure
times, enough images for a thorough cosmic-ray rejection and a
satisfactory delay between the `reference' and `discovery' images, we
have none the less discovered a candidate event in the rich cluster
AC\,114. The core of cluster AC\,114 ($z=0.31$) was observed on three
occasions with WFPC-2 in the F702W filter (see Natarajan et al.~1998).
The first two observations have total exposure times of 16.8\,ks each
and were taken on JD\,50086.3 and 50089.3 (GO\#5935). Both of these
observations suffered from scheduling problems which resulted in the
images having $\sim 4$--5$\times$ the nominal sky background, limiting
their sensitivity to $R_{702}\sim 26$ rather than the expected
$R_{702} \sim 27$.  As a result the cluster core was re-observed in
Director's discretionary time (GO\#7201) on JD\,50747.5 for a further
15.6\,ks.

Each of the observations comprises 6 single-orbit exposures in 3
pairs, each of which was spatially offset by 3 WFC pixels to enable
the removal of hot pixels and similar artefacts.  The observations are
shifted so as to place the cluster cD on WFC4 (JD\,50086.3), WFC2
(JD\,50089.3) and on the boundary of WFC3--WFC4 (JD\,50747.5) with
space-craft orientations of PA(V3)=270, 270 and 220 degrees
respectively.  The three observations of the cluster core therefore
sample very different areas of the WFPC-2 detectors, and the
combination of this with the staggered timing makes this dataset well
suited for searching for faint variable sources.

The data was reduced using the standard STScI pipeline, aligned with
integer pixel shifts and then stacked using the {\sc stsdas} task {\sc
  crrej}.  The observations were calibrated using the F702W zero-point
quoted by Holtzman~et~al.~\shortcite{holtzman95}.  Finally, the three
observations were rotated and aligned to allow a search for objects
whose brightness varies between the visits.

\begin{figure*}
\epsfig{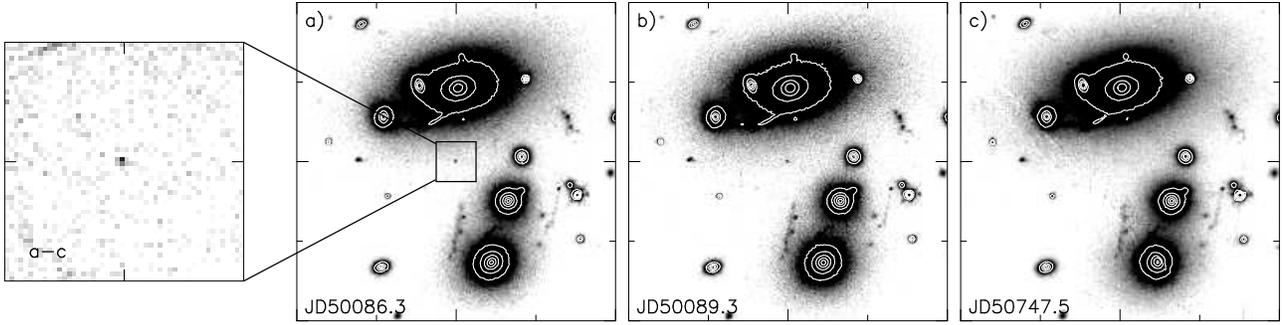}
\caption{A demonstration of the technique applied to \hst\ archival
  data on the lensing cluster AC\,114 ($z=0.31$). The detection
  $R_{702}=26.3$ is found in both reference images on different parts
  of the detector (and hence cannot be an instrumental defect), but it
  is invisible ($R_{702}>27.6$) in the image taken 21 months later.}
\label{ac114}
\end{figure*}

The overlapping area between the three observations comprises a total
area of roughly $80''\times 80''$ and this region was searched for
sources which were seen in both the JD\,50086.3 and 50089.3
observations, but which were missing from the JD\,50747.5 observation.
The sequence is shown in Fig.~\ref{ac114}.  A variable point-source
was identified -- 4.6$''$ (45\,kpc) out in the halo of
the central galaxy.  The source brightness is $R_{702}=26.4\pm 0.15$
and $26.2\pm 0.15$ in the earlier observations, but is fainter than
$R_{702}>27.6$ ($3\sigma$) in the observations taken 21 months later.

Though the object could be a SNe\,Ia in the envelope of the central cD
galaxy, this would make it very faint unless caught at the end of its
light curve. Note that the absence of a counterimage is not unexpected
given the likely time delay involved.

Gal-Yam \& Maoz~(2000) have conducted a similar comparison of the
strong-lensing cluster A\,2218 which revealed another potential SNe.
They detect a transient object ($V_{606W}=22.8$) in an F606W March
1999 image, but which is absent from both an F606W January 2000 and an
F702W September 1994 image.  Though the relative brightness of this
SNe suggests that it occurred in a $z=0.1753$ cluster member galaxy,
the usefulness of strategically surveying strong lensing clusters, for
both lensed and unlensed SNe, is well demonstrated.

\section{Discussion}
\label{discussion}

As we have seen in $\S$\ref{predictedcounts}, there is a high degree
of likelihood of detecting a new SNe in a repeat visit to a strong
lensing cluster. We will now discuss possible follow-up strategies to
any detections, and the type of science that can be achieved from
these types of surveys.

As with early ground-based SNe\,Ia searches conducted in rich clusters
\cite{hansen89}, it could be difficult to isolate the SNe, constrain
its type, and determine a redshift. These early SNe searches were,
however, essential in motivating astronomers to execute the later more
ambitious programs, and before any concerted follow-up programs can be
developed, it is first necessary to determine how many lensed SNe
there may be and, where possible, their properties.

One of the primary benefits of the lensed search is the fact that
faint, high-$z$ SNe -- which would otherwise be too faint for
spectroscopic follow-up -- are brightened sufficiently for
spectroscopy to become feasible. We estimate that $\simeq40$ per cent
of $z>1.5$ sources could be followed in this way, compared to
$\simeq5$ per cent without lensing. Such highly-lensed SNe such as
these will typically be identifiable via their proximity to the
cluster core. However, the magnitude of the faintest detected objects
is still too faint for spectroscopic follow-ups on existing equipment.
We need to examine various other methods of exploiting discoveries of
lensed high-$z$ SNe, and in particular, the possibilities of
distinguishing the type of SNe in order to preferentially select
either SNe\,Ia or II.

\begin{figure}
\epsfig{figure=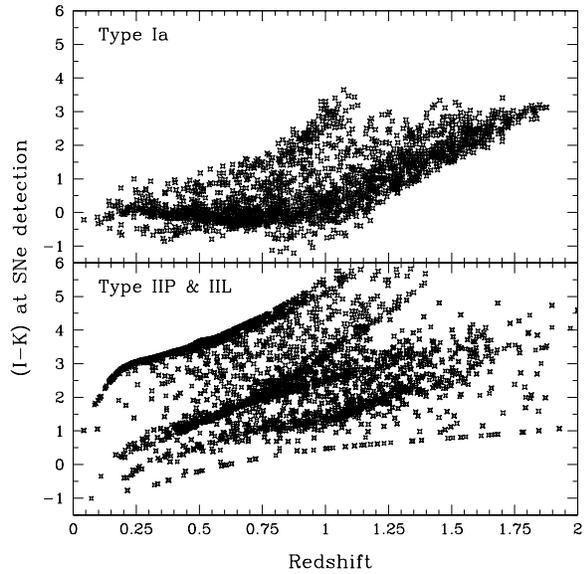,width=80mm}
\caption{The simulated $I-K$ colour-redshift distribution for likely
  detections of SNe in a forward search behind a $z=0.2$ rich cluster
  to $I_{814}\le26.0$. Each point represents a different detection,
  taking account of visibility, $k$-corrections, dust (SNe\,II only),
  and lens magnification effects. The lensing assumes the NFW profile
  discussed earlier. The distribution indicates the feasibility of
  ground-based follow-ups.}
\label{magz}
\end{figure}

New SNe will be caught at various stages on their light curve. We
consider a statistical approach to classifying the detected SNe based
on their colour distribution using a ground-based follow-up to any
$I_{814}$ detections using, as an example, the NIRC on the
Keck-II telescope.  Fig.~\ref{magz} shows the $(I-K)$ --
redshift distribution of detected SNe in the \hst\ search using a
simulation of many repeat visits to a $z=0.2$ cluster, with the search
criteria described in $\S$\ref{predictedcounts}. The `ridgeline' of
blue SNe\,Ia colours arises from those nearer maximum light, with the
SNe\,IIs redder than the Ia's.  The ridgeline of redder SNe\,IIs
arises from the fact that SNe\,IIs settle down to a constant colour
after $\simeq 50$ days in the restframe. The other ridges and gaps in
the diagram are a product of the light curve, for example there is a
greater likelihood of detecting a SNe in a plateau phase than during a
steep decline.

\begin{figure}
\epsfig{figure=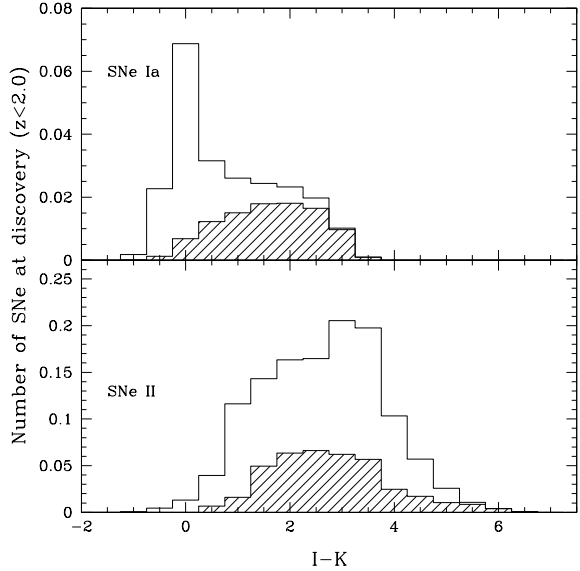,width=80mm}
\caption{The distribution in $I-K$ of detected SNe for both Ia and II types.
  The shaded regions indicate the colours of the highest redshift
  objects ($z>1$) and hence how follow-up strategies designed only
  to select high-$z$ candidates might be imagined.}
\label{magnimk}
\end{figure}

Fig.~\ref{magnimk} indicates that the bulk of the reddest objects
would be expected to be SNe with $z>1$. These red objects would be
detected in follow-ups using large ground-based telescopes in the
$K$-band; i.e. those detected in $K$ are likely to be higher-$z$ and
likely to be SNe\,II. Several photometric measurements can determine
the rate of decline over a lunation and further constrain the type.
The bluest detections will be SNe\,Ia observed close to maximum light.
These may be further distinguished via the presence of a host galaxy
whose redshift can be measured.

Although simply determining the number counts in searches such as
these is important, it is interesting to consider other applications
and their impact on our understanding of SNe.

The number counts of SNe\,Ia predicted is small, and it would be
difficult to advance the existing Hubble diagram using this method.
Accurate amplification corrections to the detected magnitudes would
also be required. The mass models of most rich clusters are currently
accurate to around 10--20 per cent, which would lead to uncertainties
in the magnitudes of around 0.5\,mag, insufficient for distinctions of
a cosmological world model to be made. However, the \textit{mere
  detection} of a high-$z$ (and therefore red) SNe\,Ia would have
important implications for the metallicity of the progenitor system.
In a survey conducted to $I=26.0\,(27.0)$ across 40 clusters, we would
expect $5(8)$ SNe\,Ia to be found in the forward search, 2--3 of which
would be at $z>1$. Using these, the first tests of the high-$z$
SNe\,Ia rate could begin.

A significant achievement would be the determination of the high-$z$
SNe\,II rate, no measurement of which yet exists. The SNe\,Ia rate
study of Pain et al.~\shortcite{pain00a} at $z=0.55$ shows that a
sample of 38~SNe results in statistical and systematic uncertainties
in the SNe rate of $\simeq 20$ per cent. Again, in a survey of 40
clusters, we would expect around $35(60)$ SNe\,II to be detected in
the `forward' search which could then be followed up from the ground.
Of these $\simeq10(17)$ would be SNe\,II at $z>1$, and using the
cluster mass models and an assumed cosmology, measures of the SFR at
$z=1$--2 could be made in an independent manner to existing
measurements.

All these rates will be increased by using a larger field-of-view
instrument, for example the 11.33\,$\rmn{arcmin}^2$ Advanced Camera
for Surveys (ACS). As an example, we repeated the simulations using
this instrument and the $\tau=0.3\,\rmn{Gyr}$, SFH-II, $I=26.0$ survey
limit scenarios, shown as the bottom line in Table~\ref{count_table}.
With this instrument, the predicted counts to $I=26.0$ are as high as
using the WFPC-2 to a survey limit of $I=27.0$, and it becomes almost
certain that a SNe will be found, and the larger field-of-view, and
hence coverage of the lensing cluster, results in the potential
discoverey of more high-$z$ SNe.

\section{Conclusions}

In this paper, we have presented a technique for detecting SNe at
$z>1$ by utilising the magnification bias obtained by searching
through rich clusters of galaxies. We summarise our main conclusions
as follows:

\begin{enumerate}
  
\item{} We have shown that a search utilising repeat \hst\ WFPC-2
  visits covering the cores of rich $z=0.2$--0.6 clusters would reveal
  $\simeq1$--3 new SNe, of which $\simeq80$ per cent will be of type
  II. Such a search will both increase the mean redshift of detected
  SNe, and will result in a threefold increase in the number of
  $z>1.5$ SNe detected.
\item{} We have applied our technique using \hst\ archival
  data, and have discovered a possible SNe event in the cluster AC\,114 at
  $z=0.31$.
\item{} We have discussed various follow-up strategies. The brightest
  events (to $I=24$) can be followed spectroscopically using large
  ground-based telescopes, and a fraction of these will be highly
  lensed $z>1.5$ SNe, which it would not be possible to follow-up
  without the lensing boost. For fainter SNe, $K$-band imaging will
  detect and select the redder, high-$z$ SNe, and the light curve
  measurements can constrain the type.
\item{} We have discussed the potential scientific benefits
  of such a search, including the determination of the high-$z$ SNe
  rates, constraints on the cosmic SFH, and the possibility of
  eliminating some SNe\,Ia progenitor models.
\item{} We briefly consider the merits of using the Advanced Camera
  for Surveys, and find that even to $I=26.0$, we expect $\simeq3$ SNe
  per search (one of which will lie at $z>1$) with a greater extension
  in the redshift range probed.

\end{enumerate}

\section*{Note added in proof}

We note the preprint of Saini et al.~(2000), which also examines the
effect of gravitational lensing by massive clusters on high-redshift
SNe, in particular those located near to giant arcs where the lensing
effect is strongest. For the case of the cluster A\,2218, they
conclude such arcs could enhance the S/N of detected SNe by
$\simeq15$, and that such lensing would substantially extend the
maximum redshift of SNe detection. We note that they also advocate the
monitoring of such systems for the detection of cosmologically useful
SNe at $z\simeq1$ and beyond.

\section*{Acknowledgments}

We thank the anonymous referee whose comments improved this
manuscript. We also thank Reynald Pain for providing his
pre-publication value of the SNe\,Ia rate, and Saul Perlmutter,
Jean-Paul Kneib and Cristiano Porciani for their many helpful
discussions in preparing this paper. This research used resources of
the National Energy Research Scientific Computing Center, which is
supported by the Office of Science of the U.S. Department of Energy
under Contract No. DE-AC03-76SF00098.

\label{lastpage}

\end{document}